\documentclass[12pt]{article}

\usepackage{epsf}
\usepackage{amsmath}
\usepackage{graphics}
\usepackage{amsfonts}
\usepackage{amssymb}
\usepackage{latexsym}
\usepackage{color}
\usepackage[dvips]{graphicx}
\input{colordvi.tex}

\newcommand{\nn}{\nonumber \\}

\begin{document}

\begin{titlepage}

\begin{flushright}
UT-11-25\\
IPMU 11-0135
\end{flushright}

\vskip 3cm

\begin{center}

{\large \bf
NMSSM in gauge-mediated SUSY breaking \\without domain wall problem
}
\vspace{1cm}

{Koichi Hamaguchi$^{(a,b)}$, Kazunori Nakayama$^{(a)}$ and Norimi Yokozaki$^{(a)}$}

\vskip 1.0cm

{\it
$^a$Department of Physics, University of Tokyo, Bunkyo-ku, Tokyo 113-0033, Japan\\
$^b$Institute for the Physics and Mathematics of the Universe,
University of Tokyo, Kashiwa 277-8568, Japan\\
}

\vskip 1.0cm

\begin{abstract}
A problem of the gauge-mediated SUSY breaking model is its difficulty to generate a natural value of
$\mu/B\mu$, while the NMSSM is a natural framework to solve the $\mu/B\mu$ problem.
The NMSSM in gauge-mediated SUSY breaking in its original form does not work well
since the singlet field cannot develop a desired vacuum expectation value.
It also suffers from the cosmological domain wall problem.
We study an extension of the model to include additional vector-like matter,
which is charged under the hidden QCD.
It is shown that this simple extension solves both the problems.
We study phenomenological and cosmological implications of this extended models.
The lightest Higgs mass can be as large as 130--140 GeV for some model points.
\end{abstract}

\end{center}
\end{titlepage}

\section{Introduction}

While the minimal supersymmetric standard model (MSSM)
is well motivated as a physics beyond the standard model from the viewpoint of the gauge hierarchy problem,
it has a huge number of parameters in its general form once the SUSY breaking effects are taken into account.
In order to suppress the unwanted flavor changing and CP-violating processes,
these SUSY breaking parameters must be controlled carefully.
Patterns of SUSY breaking parameters are not determined unless the mechanism of SUSY breaking is specified.

Gauge-mediated SUSY breaking (GMSB)~\cite{Giudice:1998bp} models provide a beautiful framework.
In the GMSB model, the SUSY breaking effect is transmitted to the visible sector by the gauge interaction,
and hence the SUSY breaking parameters are induced in such a way that the SUSY flavor and CP problems are significantly relaxed.\footnote{Even in GMSB models, a GUT breaking operator and supergravity effects can induce sizable CP violating effects in general \cite{Moroi:2011fi}. However, the supergravity effects are negligible if the gravitino mass is sufficiently light, and GUT breaking effects depend on how the GUT is broken. Therefore we do not consider those effects in this study.
}
On the other hand, GMSB models suffer from a so-called $\mu / B\mu$-problem.
In the MSSM there is one supersymmetric dimensionful parameter, $\mu$, which appears in the superpotential as
$W = \mu H_u H_d$ where $H_u$ and $H_d$ denote up- and down-type Higgs superfields.
The SUSY breaking effect generally induces the scalar potential as
$V = B\mu H_u H_d + {\rm h.c.}$.
Both $\mu$ and $B$ must be around the weak scale in order to have a correct electroweak (EW)
symmetry breaking vacuum. 
At first sight, however, there seems to be no reason why it is so. This is the $\mu/B\mu$-problem.

In order to solve the $\mu$-problem, we first need to introduce some symmetry which forbids 
the $\mu$-term in the exact symmetry limit, and generates sizable $\mu$ value
as result of (either explicit or spontaneous) breaking of the symmetry.
A simple idea is to replace the $\mu$ with a singlet field $S$, 
which is charged under the symmetry, as
\begin{equation}
	W = \frac{\lambda}{M_P^{n-1}} S^n H_u H_d,  \label{SHuHd}
\end{equation}
where $M_P$ is the reduced Planck scale.
The $\mu$-parameter is dynamically generated by the vacuum expectation value (VEV) of $S$ :
$\mu = \lambda \langle S \rangle^n/M_P^{n-1}$.
The Peccei-Quinn (PQ) symmetry, U(1)$_{\rm PQ}$, is one of the candidates.
These fields are assumed to have charges of $S(+1), H_u(-n/2), H_d(-n/2)$ under the U(1)$_{\rm PQ}$.
All other terms involving $S$ are forbidden due to the U(1)$_{\rm PQ}$.
If the U(1)$_{\rm PQ}$ is a good symmetry, the almost massless pseudo Nambu-Goldstone boson, axion,
appears in association with the spontaneous breakdown of the symmetry, and the physics of the axion
constrains the VEV of $S$ as $10^9{\rm GeV}\lesssim \langle S \rangle \lesssim 10^{12}$GeV.
It can give rise to a sizable $\mu$-term for $n=2$~\cite{Kim:1983dt}.
The $\mu/B\mu$-problem in the framework of 	PQ symmetric GMSB model
has been recently investigated in Ref.~\cite{Jeong:2011xu}.

Instead of the PQ symmetry, the discrete symmetry, $Z_3$, is sufficient to forbid all dimensionful couplings.
The allowed terms are
\begin{equation}
	W = \lambda S H_u H_d + \frac{\kappa}{3}S^3.
\end{equation}
It is then easy to see that $S$ is stabilized at TeV scale, if the SUSY breaking mass
term for the $S$ is also of the order of TeV scale.
This class of models is called next-to-MSSM (NMSSM)~\cite{Ellwanger:2009dp}.
It is nontrivial, however, whether this mechanism works in GMSB models
since $S$ is a gauge singlet and its SUSY breaking mass must be suppressed.
Then it becomes difficult to have correct EW symmetry breaking vacuum.
Moreover, since $Z_3$ is spontaneously broken, domain walls (DWs) are formed in the early Universe.
DWs eventually dominate the energy density of the Universe and they change
the standard cosmological evolution scenario drastically.
One might introduce an explicit $Z_3$ breaking term by hand as a possible solution
to the DW problem, 
but it was pointed out in Ref.~\cite{Jain:1994tk,Abel:1995wk,Abel:1996cr} that 
such a term generates a large tadpole term for the singlet $S$ and reintroduces the hierarchy problem.\footnote{
	Sophisticated choices of the discrete symmetry, rather than the $Z_3$, might allow us to 
	have a moderate tadpole term~\cite{Panagiotakopoulos:1998yw}.
	See also footnote \ref{footnote:10eV}.
}

Both of these problems are solved if we introduce additional vector-like matter 
having the $Z_3$ and QCD color charge, $Q$ and $\bar Q$, which couple to $S$ as
\begin{equation}
	W = \lambda_Q' S Q' \bar Q' .  \label{SQQ}
\end{equation}
The direct coupling of $S$ to $Q' (\bar Q')$ significantly affects the soft mass of $S$
through the renormalization group evolution,
and may lead to correct EW symmetry breaking vacuum.
Also the existence of $Q$ and $\bar Q$ makes the $Z_3$ anomalous under the QCD at the quantum level.
Thus $Z_3$ is not an exact symmetry and the DW is unstable~\cite{Preskill:1991kd},
solving the cosmological DW problem.
This simple extension solves potentially harmful problems in the original NMSSM in GMSB.
Actually this kind of models was studied in literatures~\cite{Dine:1993yw,Agashe:1997kn,deGouvea:1997cx,Liu:2008pa,Morrissey:2008gm}.\footnote{
	NMSSM in the anomaly-mediation model with an extension of vector-like matter
	was studied in Ref.~\cite{Ibe:2004gh}.
}

This solution to the DW problem, however, is not consistent with the PQ solution to the strong CP problem~\cite{Preskill:1991kd}.
This is because a non-anomalous $Z_3$-symmetry appears
by combining the original $Z_3$ and the U(1)$_{\rm PQ}$, which again makes the degenerate vacua.\footnote{
	This U(1)$_{\rm PQ}$ is different from that described above.
	MSSM fields as well as the singlet $S$ are not charged under this U(1)$_{\rm PQ}$. 
}
Thus we are tempted to make a slight modification on the model.
As a simple extension, we take the additional vector-like matter $Q'$ and $\bar Q'$ to be charged under a {\it hidden} gauge group.
If the hidden gauge coupling becomes strong at a dynamical scale $\Lambda_H$ smaller than 
the weak scale, both problems mentioned above are solved in a similar way
while maintaining the PQ solution to the strong CP problem.
Therefore, our model is free from the potential problems including
SUSY flavor and CP problems, $\mu/B\mu$ problem, DW problem and 
also compatible with the PQ solution to the strong CP problem.

In the next section we briefly review the NMSSM in GMSB with vector-like exotics,
and then in Sec.~\ref{sec:hidden}, we discuss the model with hidden vector-like exotics.

\section{NMSSM with visible vector-like matter}

\subsection{Model}

First, let us briefly review the model~\cite{Dine:1993yw,Agashe:1997kn,deGouvea:1997cx,Liu:2008pa,Morrissey:2008gm}, which includes vector-like exotics. This model has following
 superpotential,
\begin{equation}
	W = W_{\rm MSSM}+ \lambda S H_u H_d + \frac{\kappa}{3}S^3 + S \left (\lambda_{D'} D'\bar D'
	 + \lambda_{L'} {L'}\bar L'\right ),
\end{equation}
and the corresponding soft terms,
\begin{eqnarray}
	-{\cal L}_{\rm soft} &=&  
	-{\cal L}_{\rm soft}^{\rm MSSM} +
	\lambda A_\lambda S H_u H_d + 
	\frac{\kappa}{3} A_\kappa S^3 + 
	S \left (\lambda_{D'} A_{\lambda_{D'}} \tilde{D'}\tilde{\bar D}'
	 + \lambda_{L'} A_{\lambda_{L'}} \tilde{L'}\tilde{\bar L}' \right ),
	\nonumber\\
	&& 
	+ m_{D'}^2 |\tilde{D}'|^2 
	+ m_{\bar{D}'}^2|\tilde{\bar{D}}'|^2
	+ m_{L'}^2 |\tilde{L}'|^2 
	+ m_{\bar{L}'}^2|\tilde{\bar{L}}'|^2
        + m_S^2 |S|^2
        \label{eq:Lsoft}
\end{eqnarray}
where 
$D'$ $(\bar D')$ are ${\bf 3}$ (${\bf\bar 3}$) representations of the SU(3)$_C$
and SU(2)$_L$ doublets $L'$ and $\bar L'$ are introduced in order to maintain the gauge coupling unification.
The vector-like matters  obtain masses of $\lambda' \langle S \rangle$. The scalar components also receive SUSY breaking masses.

In the original NMSSM without vector-like matters, viable sparticle masses cannot be obtained with soft SUSY breaking parameters generated by GMSB: 
Higgs and sparticles are unacceptably light in order to satisfy the stationary conditions 
for $v_u (\equiv \langle H_u \rangle)$, $v_d (\equiv \langle H_d \rangle)$ 
and $v_s (\equiv \langle S \rangle)$~\cite{deGouvea:1997cx}. 
The reason is that the soft masses should be small to satisfy the stationary conditions, 
due to the smallness of effective $\mu$ term, $\lambda v_S$.
In order to raise the particle masses, we need a sizable value of $v_S$, 
which is induced by negative $m_S^2$ and/or large trilinear couplings $A_\lambda$ and $A_\kappa$.
Although it is difficult to realize such a situation in the original NMSSM matter content, 
adding extra-vector like quarks  can lead to a large negative value of $m_S^2$, resulting in sufficiently large $v_S$.
This is seen in the renormalization group equation (RGE) for $m_S^2$,
\begin{eqnarray}
	\frac{d m_S^2}{dt} &=& \left(\frac{d m_S^2}{dt}\right)_{\rm NMSSM} 
	+ \frac{2}{16\pi^2}\left[ 3 \lambda_{D'}^2 \sum_{i=1}^{N_F} (m_{D'}^2 
	+ m_{\bar{D'}}^2 + |A_{\lambda_{D'}}|^2 + m_S^2   ) \right]
	 \nn
	&+& \frac{2}{16\pi^2}\left[ 2 \lambda_{L'}^2 \sum_{i=1}^{N_F} (m_{L'}^2 
	+ m_{\bar{L'}}^2 + |A_{\lambda_{L'}}|^2 + m_S^2   ) \right],
	\label{eq:ms_rge}
\end{eqnarray}
where we have introduced $N_F$ vector-like matters,
$t=\ln Q$ and the first term is the RGE without the vector-like matters, which is given by
\begin{eqnarray}
	\left(\frac{d m_S^2}{dt}\right)_{\rm NMSSM} = \frac{2}{16\pi^2} 
	\left[ 2\lambda^2 (m_{H_d}^2 + m_{H_u}^2 + m_S^2 + |A_\lambda|^2) + 2 \kappa^2 (3 m_S^2 + |A_\kappa|^2)\right] .
\end{eqnarray}
This term contains large negative contributions from $m_{H_u}^2$ near the electroweak scale, 
which prevents $m_S^2$ from having a sufficiently large negative value at the weak scale. 
The second and third terms in Eq.~(\ref{eq:ms_rge}) denote the contributions from vector-like quarks $D'$ and leptons $L'$, respectively. 
These additional contributions can lead to sufficiently large negative $m_S^2$ when $m_{D'}^2$ 
and $m_{L'}^2$ are large, in analogy that $m_{H_u}^2$ is driven to be negative by the stop contributions.

\subsection{Cosmological issues}

Now let us turn to cosmological issues of this model.
If the vector-like matters do not mix with the MSSM particles, 
the lightest one among them is stable and may be a candidate of dark matter (DM) in the Universe.
Actually the mixing between vector-like matters and MSSM matters are forbidden by 
assigning an additional parity or U(1) symmetry, under which only the vector-like matters transform.
However, none of them is allowed as a dominant component of DM once we take account of
stringent constraints on relic strongly-interacting and electrically charged particles.
Even the electrically neutral component of $L'$ is excluded as a dominant component of DM,
because it has too large scattering cross section to a nucleon
through the coherent $Z$-boson exchange~\cite{Srednicki:1986vj,Falk:1994es}
which exceeds the current limits from the DM direct detection experiments~\cite{Aprile:2011hi}.
To remedy this, we may introduce small mixings between additional vector-like matter and SM particles
in order to make the vector-like matters unstable.
Although large mixings are prohibited by the constraints from flavor-changing processes,
even a small mixing makes the lifetime of the vector-like matter sufficiently short 
so that it decays much before big-bang nucleosynthesis (BBN).
Therefore, vector-like matters do not contribute to the energy density of the Universe.
Instead, the gravitino may be a DM candidate for appropriate 
reheating temperature after inflation~\cite{Moroi:1993mb}.

One of the serious problems in the NMSSM is the DW problem~\cite{Vilenkin:1981zs}.
Around the epoch of EW phase transition, 
$S$ obtains a VEV and $Z_3$ symmetry is spontaneously broken.
Then the DW is formed whose tension, $\sigma$, is of the order of $\sim (1{\rm TeV})^3$.
Denoting by $R(T)$ the typical scale of irregularity on the DW at the cosmic temperature $T$, 
the DW energy density is estimated as $\rho_{\rm DW}(T) \sim \sigma / R(T)$.
As long as the friction due to the thermal plasma is efficient, $R(T)$ is given by~\cite{Preskill:1991kd}
\begin{equation}
	R(T) \sim \frac{\sqrt{\sigma M_P}}{T^3}.
\end{equation}
Irregularities smaller than the scale of $R(T)$ is smoothed out by the DW tension.
This becomes equal to the Hubble radius at $T\sim \sqrt{\sigma/M_P}$
and at the same time it begins to dominate the Universe.
As long as the $Z_3$ symmetry is exact, the DW is stable and it causes serious cosmological problems.

In the present model, however, $Z_3$ is not an exact symmetry at the quantum level.
This is because the $Z_3$ rotation involves chiral rotations of quarks, which is anomalous under the QCD.
Thus the effect of quantum anomaly violates the $Z_3$ symmetry~\cite{Preskill:1991kd,Abel:1995wk}.
Without the additional quarks $D'$ and $\bar D'$, 
$Z_3$ symmetry is not violated because there exist three generations of quarks.
Thus we need additional quarks $D'$ and $\bar D'$ charged under SU(3)$_C$
in order to make the $Z_3$ anomalous.\footnote{
	Introduction of one vector-like quark pair is sufficient for this purpose.
	If three such quark pairs are introduced, $Z_3$ again becomes non-anomalous.
}
Since the $Z_3$ has an anomaly under the QCD, 
the degeneracy among three discrete vacua are lifted completely.
The magnitude of the bias in the scalar potential is estimated to be $V_\epsilon \sim \Lambda_{\rm QCD}^4$,
where $\Lambda_{\rm QCD}\sim 200$MeV is the QCD scale.
This effect is turned on after the QCD phase transition.
It serves as a bias for the DWs to collapse and disappear~\cite{Moroi:2011be}.


\subsection{Compatibility with the Peccei--Quinn mechanism}
\label{sec:DWandPQ}

A shortcoming of this model is that it is incompatible with the PQ mechanism~\cite{Peccei:1977hh}
for solving the strong CP problem, as shown in Ref.~\cite{Preskill:1991kd}.
A crucial point is that both $Z_3$ and U(1)$_{\rm PQ}$ have anomaly under the same QCD
and they mix to form another unbroken $Z_3'$ symmetry.
Under the new $Z_3'$ symmetry, DWs are stable and harmful.\footnote{
	This U(1)$_{\rm PQ}$ cannot be same as that described in the Introduction
	because of the term in the superpotential, $S^3$.
	An example of required U(1)$_{\rm PQ}$ symmetry will be given in Sec.~\ref{sec:PQ}.
}
Thus we need to rely on another mechanism in order to solve the strong CP problem~\cite{Nelson:1983zb}.
In the next section we construct a variant model of the NMSSM in GMSB with hidden exotics
which is fully compatible with the PQ solution to the strong CP problem.\footnote{
Another possible solution is a low scale gauge mediation with an ultralight gravitino,
supplemented by an explicit $Z_3$ breaking operator suppressed by Planck scale.
The induced tadpole operators, $\delta V \sim (S M_{P} m_{3/2}^2 + F_S M_{P} m_{3/2} + {\rm h.c.})/(16\pi^2)^n$
\cite{Abel:1996cr} can be large enough to break $Z_3$, while not destabilizing the EW scale, 
if the gravitino mass $m_{3/2}$ is sufficiently small.
\label{footnote:10eV}
}

\section{NMSSM with hidden vector-like matter}
\label{sec:hidden}

\subsection{Model and mass spectrum}   \label{sec:mass}

We have encountered a difficulty in the previous model for solving the strong CP problem while making DWs unstable.
These problems do not exist if we take the additional vector-like matter to be charged under a {\it hidden} QCD
but not under SU(3)$_C$. 
Let us consider the following superpotential,\footnote
{We neglect the flavor dependence of $\lambda_{Q'}$ for simplicity. However including the flavor dependence does not change the result qualitatively.}
\begin{equation}
	W = W_{\rm MSSM}+ \lambda S H_u H_d + \frac{\kappa}{3}S^3 + \lambda_Q' S Q'_i\bar Q'_i,
	\label{super}
\end{equation}
and the corresponding soft terms similar to Eq.(\ref{eq:Lsoft}).
Here, $Q'$ and $\bar Q'$ are fundamental and anti-fundamental representations of the 
hidden QCD : SU$(N)_H$.
For concreteness, we consider the following messenger and SUSY breaking sector~\cite{Murayama:2006yf}
\begin{equation}
	W_{\rm mess} = X( k \Psi \bar \Psi + k' \Psi' \bar \Psi' - M^2 ) 
	+ m_\Psi \Psi \bar \Psi  + m_{\Psi'} \Psi' \bar \Psi' , \label{eq:mess}
\end{equation}
where $\Psi$ and $\bar\Psi$ are messengers giving rise to soft terms  for the MSSM particles
and they are assumed to be fundamental- and anti-fundamental representations of SU$(5)$.
$\Psi'$ and $\bar\Psi' $ are the ``hidden'' messenger fields which give rise to the SUSY breaking mass 
for the scalar components of $Q'$ and $\bar Q'$.\footnote{
The K\"ahler potential is assumed to be of the form of
$K = |X|^2 - |X|^4/4\Lambda^2.$
Then the $X$ obtains a positive mass around the origin.
There is a SUSY breaking metastable vacuum at $X=\Psi=\bar\Psi=\Psi'=\bar\Psi'=0$ if $m_\Psi^2 > kM^2$
and $m_\Psi'^2 > k'M^2$.
	SUSY breaking vacuum exists even if there are not bare messenger mass terms
	by taking into account the supergravity effect~\cite{Kitano:2006wz}.
}
Charge assignments on the fields are summarized in Table
  1.\footnote{The
$Z_3$ symmetry can be compatible with the 
neutrino mass term in the superpotential
$W=(LH_u)^2/M_N$ where $M_N$ is the seesaw scale~\cite{Gogoladze:2008wz}.}

The soft masses of the hidden gauginos and sfermions at the messenger scale are given by
\begin{eqnarray}
	M_{\tilde g'} &=& N_H \frac{g_H^2}{16\pi^2} \Lambda_{\rm mess}, \\
	m_{Q'_i}^2 &=& m_{\bar{Q'_i}}^2= 2N_H C_H \frac{g_H^4}{(16\pi^2)^2} 
	\Lambda_{\rm mess}^2,
\end{eqnarray}
where $\Lambda_{\rm mess}\equiv k'M^2/m_{\Psi'}$,
$g_H$ is the hidden gauge coupling constant, $N_H$ is the number of hidden messengers
and $C_H=(N^2-1)/2N$ is the quadratic Casimir invariant for the gauge group SU$(N)_H$.


\begin{table}[t!]
  \begin{center}
    \begin{tabular}{ | c | c | c | c | c | c | c | c | c | c | c |}
      \hline 
         ~          &  $S$  & $H_u$ & $H_d$ & $Q'$ & $\bar Q'$ &$\Psi_c$ & $\bar \Psi_c$&$\Psi'$ & $\bar \Psi'$ & $X$\\
       \hline \hline
         $Z_3$  & $1$  & $1$ & $1$  & $1$   & $1$   & $0$ & $0$ & $0$ & $0$ & $0$     \\ \hline 
       SU(3)$_C$ &${\bf 1}$& ${\bf 1}$ & ${\bf 1}$  & ${\bf 1}$ & ${\bf 1}$ & ${\bf 3}$ & ${\bf \bar 3}$ & ${\bf 1}$ & ${\bf 1}$ & ${\bf 1}$  \\ \hline
       SU$(N)_H$ & ${\bf 1}$ & ${\bf 1}$& ${\bf 1}$ & ${\bf N}$ & ${\bf \bar N}$ & ${\bf 1}$ & ${\bf 1}$ & ${\bf N}$ & ${\bf \bar N}$  & ${\bf 1}$ \\ \hline   
    \end{tabular}
    \caption{ 		
     	Charge assignments on chiral superfields fields in the model under 
	the $Z_3$, SU(3)$_C$ and SU$(N)_H$.
	$\Psi_c$ denotes the colored component of the messenger field.
     }
  \end{center}
  \label{table:charge}
\end{table}

In analogy with the model with visible vector-like matter, successful EW symmetry breaking
is achieved by the loop effect from the hidden vector-like squarks, hence we do not repeat the discussion.
Note that there is no SUSY CP problem in this model.
If $\lambda_Q'$ is too large, the Yukawa couplings blow up quickly. 
This can be seen from the beta-functions for the corresponding Yukawa couplings:
\begin{eqnarray}
	\frac{d \lambda}{dt} &=&  \left(\frac{d\lambda}{dt} \right)_{\rm NMSSM} 
	+ \frac{1}{16\pi^2} 3 \lambda N_F \lambda_{Q}'^2 \nn
	\frac{d Y_t}{dt} &=& \frac{1}{16\pi^2} Y_t \lambda^2  + \dots,
\end{eqnarray}
where $Y_t$ is the top Yukawa coupling and  the dots denote the beta-function of the MSSM. 
Therefore the value of $\lambda_{Q}'$ is bounded from above in order for the perturbativity to be maintained 
up to, say, GUT scale.

Let us discuss the vacuum stability of this model. 
The mass terms for hidden squarks ${\tilde{Q}}'$ and ${\tilde{\bar{Q}}}'$ 
around the realistic vacuum with $Q'=0$ are given by
\begin{eqnarray}
	-\mathcal{L} \ni 
	\left({\tilde{Q}}' \ \, {\tilde{\bar{Q}}'}^* \right) M_{Q'}^2 \left(
	\begin{array}{c}
	{\tilde{Q'}}^* \\
	{\tilde{{\bar{Q}}}}'
	\end{array}
	\right)
\end{eqnarray}
where
\begin{eqnarray}
	&&(M_{Q'}^2)_{11}= m_{{\tilde{Q}}'}^2 + \lambda_{Q'}^2 v_S^2, \  (M_{Q'}^2)_{22}= m_{{\tilde{\bar{Q}}}'}^2 + 	\lambda_{Q'}^2 v_S^2 ,\nn
	&&(M_{Q'}^2)_{12}= (M_{Q'}^2)_{21} =\lambda_{Q'}
	 (-\lambda  v_u v_d  + \kappa  v_S^2 + A_{\lambda_{Q'}} v_S) .
\end{eqnarray}
The term $\lambda_{Q'} \kappa  v_S^2$ may lead large mixings 
and ${\tilde{Q}}'$ may become tachyonic around this vacuum, which spontaneously breaks hidden QCD. 
Hereafter we do not consider such a case, and
we restrict ourselves to the parameters where $\tilde{Q}'$ and $\tilde{\bar Q}'$ are stabilized at the origin :
${\tilde Q}'=\tilde{\bar Q}' = 0$.

Let us discuss the {\bf NMSSM} vacuum structure modified by the inclusion of additional squarks. 
The relevant scalar potential is written as
\begin{eqnarray}
	V &=& V_F + V_D + V_{\rm soft},
\end{eqnarray}
where $V_D$ is the D-term contribution and $V_{\rm soft}$ contains soft breaking terms. 
$V_F$ is a F-term contribution, given by
\begin{eqnarray}
	V_F &=& \left|-\lambda H_u^0 H_d^0 + \kappa S^2 + \lambda_{Q'} \tilde Q'_i \tilde{\bar{Q'}}_i \right|^2 
	+ \lambda^2 |S|^2 (|H_u^0|^2 + |H_d^0|^2) \nonumber \\
	&+& \lambda_{Q'}^2 |S|^2 (|\tilde Q'_i|^2 + |\tilde{\bar Q}'_i|^2).
	\label{VF}
\end{eqnarray}
First, we investigate the vacuum structure along the direction which makes $V_D$ and $V_F$ flat. 
We take $S=0$, $|H_u^0| = |H_d^0|=v_H$ and $|\tilde Q'_i|=|\tilde{\bar{Q'}}_i|=v_Q$ to make $V_D$ flat. 
The first term in (\ref{VF}) becomes zero with 
\begin{eqnarray}
	|\lambda| v_H^2 = |\lambda_Q'| N_F v_Q^2 .
\end{eqnarray}
With this choice, 
\begin{eqnarray}
	V=V_{\rm soft} &=& 2 m_{Q'}^2 N_F v_Q^2 + (m_{H_u}^2 + m_{H_d}^2)v_H^2 \nn
	&=& \left(2 m_{Q'}^2 \left|\frac{\lambda}{\lambda_{Q'}}\right| + m_{H_u}^2 + m_{H_d}^2 \right) v_H^2.
	\label{Vflat1}
\end{eqnarray}
If this term is negative, the potential along this direction is unbounded. 
Thus the value inside the parenthesis in (\ref{Vflat1}) must be positive.
{This relation has to be carefully checked} for especially the solution with small $\lambda$.
Secondly, the potential along $H_u^0=H_d^0=0$ with vanishing first term in (\ref{VF}) 
also needs careful exploration.
The scalar potential along this direction is expressed as
\begin{eqnarray}
	V = \left(m_S^2 + 2m_{Q'}^2 \frac{|\kappa|}{|\lambda_{Q'}|}\right)v_S^2 
	+ 2\left(\frac{A_{\kappa}}{3} - A_{\lambda_{Q'}} \right)\kappa v_S^3 + 2|\lambda_{Q'} \kappa | v_S^4.
	\label{Vflat2}
\end{eqnarray}
In our analysis, we calculated the minimum along this direction and
demand that it is not deeper than the realistic EW vacuum.\footnote{
Although the potential (\ref{Vflat2}) may have a local minimum,
the required vacuum may be dynamically selected
	by thermal phase transition if the tachyonic instability develops first along the 
	direction of $|H_u|=|H_d|$ with $Q'=0$.
}
The cubic term is not relevant for realistic parameters in GMSB.

The mass spectrum and the strong coupling scale of the hidden SU(3), $\Lambda_H$, are shown in Table 2
for some typical model points. 
The calculation is performed with NMSSMTools~\cite{NMSSMTools}, which is modified to include vector-like matter (see Appendix). In the numerical analysis, we take $k=k'$ and $m_{\Psi}=m_{\Psi'}$, respectively (c.f.~Eq.(\ref{eq:mess})).
All of the points satisfy the constraints from the EW symmetry breaking and the vacuum stability. 
In these models, the messenger number is taken to be unity and 
the next-to-lightest SUSY particle (NLSP) is the neutralino. 
Generalization to models with larger messenger number is straightforward, that predict stau NLSP.
The strong scale is defined by $g_H(\Lambda_H)=4\pi$.

In P1, P2, P3 and P6, the perturbativity is maintained up to the GUT scale, 
while it is maintained up to the messenger scale in P4 and P5.
In P1, P2, P4 and P5, the lightest CP-even Higgs is SM-like, 
while in P3 and P6 the lightest CP-even Higgs is singlet-like 
and hence the LEP bound can be avoided due to the reduced coupling to the $Z$-boson. 
In P1, P2, P3 and P6, the lightest CP-odd Higgs is very light, which is a distinct property of the NMSSM. 
This CP-odd Higgs is almost singlet like. {Note that in P6, $\mu$ is positive and the experimental result of muon g-2 is explained with large $\tan\beta$.}
In the model points P4 and P5, the perturbativity is maintained only up to the messenger scale,
although the perurbative GUT unification may still be achieved if the singlet (and/or Higgs) is a composite particle (c.f.~\cite{Harnik:2003rs, Chang:2004db}).
It is interesting that the Higgs mass can be as large as 130--140 GeV in these model points,
which is difficult in usual GMSB models,
because the recent LHC data may indicate the existence of the Higgs boson around $120$--$140$GeV~\cite{EPS-ATLAS,EPS-CMS}.

\begin{table}[t!]
  \begin{center}
    \begin{tabular}{ | c | c | c | c | c | c | c | c | c | c |}
      \hline 
         ~          &  $\Lambda_{\rm mess}$  & $M_{\rm mess}$ & $g_H$ & $\lambda_{Q'}$ & $N_F$ & $\lambda$ & $\kappa$ &$\tan\beta$ & $\mu_{\rm eff}$  \\     \hline \hline
         P1  & $2\times 10^5$  & $10^6$ & $0.82$    & $0.114$   & $2$   & $0.005$ & $4.54\times10^{-4}$ & $15$ & $-764$ \\ \hline 
         P2  & $2\times 10^5$  & $10^{12}$ & $0.82$    & $0.047$   & $7$   & $0.005$ & $5.38\times10^{-4}$ & $16$ & $-959.6$ \\ \hline 
        P3 &$1.5\times 10^5$& $10^6$ & 0.74 & 0.078 & 5 & 0.005 & $4.47 \times 10^{-4}$ & 15 & $-607.6$   \\ \hline
        P4 &$ 2\times10^5$& $10^6$ & 0.96 & 0.46 & 10 & 0.7 & 0.63 & 1.5 & $-1500.8$   \\ \hline
	P5 & $1.4\times10^5$ & $10^6$ & 1.1 & 0.34 & 11 & 0.75 & 0.71 &1.4 & $-1296.0$ \\ \hline 
	P6 & $1.4\times10^5$ & $10^{10}$ & 1.0 & 0.013 & 7 & 0.005 & $-1.87\times10^{-4}$ &45 & $ 700.5$ \\ \hline 
    \end{tabular}
    \begin{tabular}{ | c | c | c | c | c | c | c | c | c | c | c |c|}
      \hline 
         ~          &  $m_{h_1}$  & $m_{h_2}$  & $m_{a_1}$ &$m_{a_2}$ & $m_{\chi_1^0}$ & $m_{\chi_1^+}$ & $\tilde{t}_1$ & $\tilde{\tau}_1$ & $\tilde{\nu}_{\tau}$  & $\tilde{g}$ & $\Lambda_H$\\
       \hline \hline
         P1  & $114.6$  & $139.6$ & $11.8$  & $1032.7$   & $140.1$   & $534.7$ & $1802.2$ & $349.1$ & $698.3$ & $1536.1$ & 0.02\\ \hline 
        P2  & $114.8$  & $205.1$ & $44.3$  & $1212.4$   & $207.5$   & $534.0$ & $1364.5$ & $494.0$ & $811.6$ & $1502.0$ & 0.02\\ \hline 
        P3 &$105.8$& $116.7$ & $9.8$ & $800.9$ & $109.5$ &$398.7$& $1371.3$ & $262.7$ & $525.6$ & $1181.2$  & $ 0.001$\\ \hline
        P4 &$132.7$& $1996.9$ & $1109.4$ & $2020.5$ & $277.8$ &$542.2$& $1715.1$ & $353.6$ & $699.1$ & $1536.5$  &0.42 \\ \hline
 P5 &$ 140.0$& $1731.2$ & $1043.9$ & $1766.3$ & $207.3$ &$407.7$& $1290.9$ & $266.1$ & $527.3$ & $1182.4$  & 1.58 \\ \hline
 P6 &$ 51.0$& $115.3$ & $ 6.29$ & $607.1$ & $52.4$ &$372.2$& $1040.2$ & $212.5$ & $526.0$ & $1092.2$  & 0.76 \\ \hline

%
    \end{tabular}
    \caption{The mass spectra of some model points are shown. The P1, P2 and P3 satisfies the constraint from perturbativity up to the GUT scale, while P4 and P5 maintain the purturbativity condition up to the messenger scale. In P6, $\mu$ is positive so that the experimental result of the muon g-2 is explained.
    All masses are written is the unit of GeV. 
     }
  \end{center}
\end{table}

\subsection{Cosmological issues}

The present model, {defined by Eq.~(\ref{super}) and (\ref{eq:mess})}, 
also has a $Z_3$ symmetry at the classical level,
and hence DWs are formed in association with the spontaneous breakdown of the $Z_3$.
Similarly to the previous model, however, the $Z_3$ symmetry has an anomaly under 
the hidden gauge group SU$(N)_H$ due to the existence of hidden quarks $Q'$ and $\bar Q'$.
If the hidden gauge becomes strong at the scale of $\Lambda_H$,
the effect of the hidden gauge instanton makes the $Z_3$ anomalous.
This introduces a bias potential which lifts the degeneracy among the three distinct vacua.
The magnitude of the bias is estimated as
\begin{equation}
	V_\epsilon \sim \Lambda_H^4.
\end{equation}
The DW energy density relative to the
bias potential at the hidden QCD phase transition, $T\sim \Lambda_H$, 
is estimated as
\begin{equation}
	\frac{\rho_{\rm DW}}{V_\epsilon} \sim \frac{1}{\Lambda_{H}}\sqrt{\frac{\sigma}{M_P}}
	\sim 2\times 10^{-5}\left( \frac{1{\rm GeV}}{\Lambda_H} \right)
	\left[ \frac{\sigma}{(1{\rm TeV})^3} \right]^{1/2}.
	\label{rho_DW}
\end{equation}
Therefore DWs disappear at the hidden QCD phase transition as long as its dynamical scale
is not much smaller than 1MeV.
In association with the collapse of DWs, gravitational waves (GWs) are emitted.
The expected energy density of GWs relative to the total energy density of the present Universe,
$\Omega_{\rm GW}$, is roughly estimated as~\cite{Moroi:2011be}
\begin{equation}
	\Omega_{\rm GW} \sim \Omega_r \frac{(\sigma/M_P)^2}{\Lambda_H^4} 
	\sim 10^{-11}\left( \frac{\sigma}{(1{\rm TeV})^3} \right)^2
	\left( \frac{1{\rm MeV}}{\Lambda_H} \right)^4,
\end{equation}
where $\Omega_r\sim 10^{-5}$ is the radiation density parameter.
The frequency extends from $\sim 10^{-12}$Hz for the lower side to 
$\sim 10^{15}$Hz for the higher side for $\Lambda_H=1$MeV
and the amplitude is flat between these frequencies~\cite{Hiramatsu:2010yz}.
The amplitude is large enough to be detected in the pulsar timing arrays
for $\Lambda_H\sim 1$ MeV, and also 
may be within the reach of future space-based gravitational wave detectors 
such as DECIGO~\cite{Seto:2001qf}.

In the present model the lightest particle in the hidden matter may be stable and relevant for cosmology.
A slight extension to include another hidden vector-like matter
charged under another hidden gauge group can easily accommodate observed DM abundance
without conflicting with the direct detection bound.
For example, we can identify the hidden gauge group as SU$(3)_H\times$SU(2)$_H$ 
where the former becomes strong at the scale of $\Lambda_H$.
We add vector-like matter $l_H (\bar l_H)$ which are (anti-)fundamental representations of SU(2)$_H$.
They are stable and annihilate through the $t$-channel SU(2)$_H$ gaugino exchange,\footnote{
	This is analogous to the WIMPless scenario~\cite{Feng:2008ya}.
}
as well as the $s$-channel singlet exchange if they have a coupling like $W = \lambda_H S l_H \bar l_H$.
A correct DM abundance may be obtained for appropriate parameter choices
and retains a beautiful WIMP scenario in the framework of GMSB.

\subsection{Compatibility with the Peccei-Quinn mechanism}  \label{sec:PQ}

{In the present case we can introduce a U(1)$_{\rm PQ}$ symmetry
so that the $Z_3$ and U(1)$_{\rm PQ}$ do not mix with each other,}
since U(1)$_{\rm PQ}$ is anomalous under the QCD while the $Z_3$ is anomalous under the hidden QCD.
Hence the DW problem is solved in the presence of the PQ symmetry, 
as opposed to the previous model.
Thus this model is compatible with the attractive PQ solution to the strong CP problem.
The simplest extension for the PQ sector is to include the following term in the superpotential
as in the KSVZ axion model~\cite{Kim:1979if} :
\begin{equation}
	W = \Phi_{\rm PQ} Q_{\rm PQ} \bar Q_{\rm PQ}, 
\end{equation}
where $\Phi_{\rm PQ}$ is the PQ field whose VEV
spontaneously breaks U(1)$_{\rm PQ}$ and $Q_{\rm PQ} (\bar Q_{\rm PQ} )$ are heavy quarks with color charge.
They have U(1)$_{\rm PQ}$ charges as $\Phi_{\rm PQ}(+2), Q_{\rm PQ} (-1), \bar Q_{\rm PQ} (-1)$
while all other NMSSM sector particles are singlet under the U(1)$_{\rm PQ}$.
{ It should be noticed that this U(1)$_{\rm PQ}$ differs from that described in the Introduction.
The stabilization of the PQ scalar at the desired VEV of $f_a=10^{10}$-$10^{12}$\,{\rm GeV}
is realized in some ways.
A simple way is to introduce the following superpotential~\cite{Banks:2002sd},
\begin{equation}
	W = \frac{\Phi_{\rm PQ}^n \bar \Phi_{\rm PQ}}{M^{n-2}},
\end{equation}
where $\bar \Phi_{\rm PQ}$ has the PQ charge $-2n$ and $M$ denotes a cutoff scale.
Then the PQ scalar is stabilized at $\langle \Phi_{\rm PQ}\rangle \sim (m_{\rm PQ}M^{n-2})^{1/(n-1)}$,
where $m_{\rm PQ}$ denotes the soft mass for the PQ scalar.
This may yield a desired value of the PQ scalar.
There may be another way to stabilize the PQ scalar~\cite{Asaka:1998ns}.
We do not specify the stabilization mechanism here, but the message is that our solution to the DW problem
in NMSSM is consistent with the PQ mechanism for solving the strong CP problem.
}\footnote{
	In the SUSY PQ model, the saxion and axino may cause cosmological problems.
	Hence the reheating temperature is strictly bounded depending on the PQ scale. 
	See e.g., Ref.~\cite{Kawasaki:2007mk}.
}

\section{Summary and discussion}

In this letter, we have studied the NMSSM in GMSB model
with inclusion of vector-like matter.
Particularly, we focused on the case that the added vector-like matters are charged under a hidden QCD.
The EW symmetry is broken successfully by the negative soft masses of the gauge singlet, 
which is induced by the loops of the vector-like squarks.
Therefore, the $\mu / B\mu$-problem, which is difficult to be explained in the GMSB models, is solved.
There is no SUSY flavor and CP problems.
The serious domain-wall problem, which is a common feature in the NMSSM, is also naturally solved
in this framework, since
the classical $Z_3$ symmetry is anomalous under the hidden QCD in the presence of vector-like matters. 
Thus domain walls are unstable and collapse at the hidden QCD phase transition.
Gravitational waves are emitted through the collapse of the domain walls, which may be detected in future experiments.
Since the anomalous $Z_3$ symmetry has nothing to do with the QCD anomaly of the PQ symmetry,
this solution to the domain wall problem is fully consistent with the PQ mechanism for solving the strong CP problem.
It is also consistent with the Higgs chaotic inflation scenario in NMSSM proposed in Ref.~\cite{Nakayama:2010sk},
which preserves the $Z_3$ symmetry and predicts observable level of primordial gravitational wave background
with tensor-to-scalar ratio of $\sim 0.05$.
The lightest Higgs mass can be as large as 130--140 GeV for some model points.



\section*{Acknowledgment}

This work is supported by Grant-in-Aid for Scientific research from
the Ministry of Education, Science, Sports, and Culture (MEXT), Japan,
No.\ 21111006 (K.N.), No.\ 22244030 (K.N.), No.\ 21740164 (K.H.), No.\ 22244021 (K.H.), and No.\ 22-7585 (N.Y.).
This work was supported by World Premier International Research Center Initiative (WPI Initiative), MEXT, Japan.

\appendix

\section{Renormalization Group Equations}

The superpotential of NMSSM + vector-like exotics is given by
\begin{eqnarray}
	W = \lambda S H_u H_d + \frac{\kappa}{3} S^3 + \lambda_{Q'_i} S Q_i' \bar{Q}_i',
\end{eqnarray} 
where $i=1 \dots N_F$.  

At the messenger scale, the soft masses for the hidden gaugino and the squarks  $Q'_i$ are induced as
\begin{eqnarray}
	M_H &=& N_H \frac{g_H^2}{16\pi^2} \Lambda_{\rm mess}, \nonumber \\
	m_{Q'_i}^2 &=& m_{\bar{Q'_i}}^2= 2  \frac{N_H}{(16\pi^2)^2} |\Lambda_{\rm mess}|^2 C_{2}^i (r) g_H^4
\end{eqnarray}
where $N_H$ is the number of hidden messengers, and $C_2^i(r)$ is the quadratic Casimir for the 
representation $r$ defined by $T^a(r)T^a(r)=C_2(r) {\bf 1}$. 
For the fundamental representation of SU$(N_C)$, it is given by $C_2(N_C)=(N_C^2-1)/(2N_C)$.

The beta function for the hidden gauge coupling is given by
\begin{eqnarray}
	\frac{d g_H}{dt} = -\frac{(3N_C-N_F)}{16\pi^2} g_H^3,
\end{eqnarray}
where $t=\ln\mu$ and $N_C$ is a number of colors, $N_F$ is a number of vector like pairs. 

The beta-function of the hidden gaugino mass is written as
\begin{eqnarray}
	\frac{d M_{H}}{dt} = -\frac{2(3N_C-N_F)}{16\pi^2} g_H^2 M_H.
\end{eqnarray}

The beta-function for $\lambda_{Q'}$ is given by the sum of the anomalous dimensions of the fields, 
which interact with the coupling:
\begin{eqnarray}
	\frac{d \lambda_{Q'_i}}{dt} = -\frac{1}{2} \lambda_{Q'_i} 
	\left(\gamma_S + \gamma_{Q'_i} + \gamma_{\bar{Q}'_i}\right),
\end{eqnarray}
where the anomalous dimensions are given by
\begin{eqnarray}
	\gamma_S &=& \frac{2}{16\pi^2}(-2\lambda^2 -2\kappa^2 
	-  N_C \sum_{i=1}^{N_F} \lambda_{Q'_i}^2), \nonumber \\
	\gamma_{Q_i} &=& \frac{2}{16\pi^2}(-\lambda_{Q'_i}^2 + 2 C_2(r) g_H^2), \nonumber \\
	\gamma_{\bar{Q}_i} &=& \frac{2}{16\pi^2}(-\lambda_{Q'_i}^2 + 2 C_2(r) g_H^2).
\end{eqnarray}
The beta-functions of $\lambda_{Q'_i}$ is explicitly written as
\begin{eqnarray}
	\frac{d \lambda_{Q'_i}}{dt} = \frac{1}{16\pi^2} \lambda_{Q'_i} 
	\left( 2 \lambda^2 + 2\kappa^2  +N_C \sum_{j=1}^{N_F} 
	\lambda_{Q'_j}^2 + 2\lambda_{Q'_i}^2 -4C_2(r) g_H^2 \right).
\end{eqnarray}
The change of $\gamma_S$ induces additional contributions to the beta function of $\lambda$, as
\begin{eqnarray}
	\frac{d \lambda}{dt} = \left(\frac{d \lambda}{dt}\right)_{\rm NMSSM} 
	+ \frac{\lambda}{16\pi^2} N_C \sum_{i=1}^{N_F} \lambda_{Q'_i}^2,
\end{eqnarray}
where the first term is given by
\begin{eqnarray}
	\left(\frac{d \lambda}{dt}\right)_{\rm NMSSM} = 
	\frac{\lambda}{16\pi^2} (4\lambda^2 + 2\kappa^2 + 3(h_t^2 + h_b^2)
	+h_{\tau}^2 -g_1^2 -3g_2^2).
\end{eqnarray}
The beta-function for $\kappa$ is also modified as
\begin{eqnarray}
	\frac{d\kappa}{dt} = \frac{\kappa}{16\pi^2}(6\lambda^2 
	+ 6\kappa^2 + 3 N_C \sum_{i=1}^{N_F} \lambda_{Q'_i}^2).
\end{eqnarray}
The RGEs of $A$-terms are given by 
\begin{eqnarray}
	\frac{d A_{Q'_i}}{dt} = \frac{2}{16\pi^2} \left( 2 \lambda^2 A_\lambda + 2\kappa^2 A_{\kappa}  
	+N_C \sum_{j=1}^{N_F} \lambda_{Q'_j} ^2 A_{\lambda_{Q'_j}} 
	+2 \lambda_{Q'_i}^2 A_{Q'_i} +4 C_2(r) g_H^2 M_H \right),
\end{eqnarray}
\begin{eqnarray}
	\frac{d A_{\lambda}}{dt} = \left(\frac{dA_{\lambda}}{dt}\right)_{\rm NMSSM}
	+ \frac{2}{16\pi^2}  N_C \sum_{j=1}^{N_F} \lambda_{Q'_j}^2 A_{Q'_j} ,
\end{eqnarray}
\begin{eqnarray}
	\frac{d A_\kappa}{dt} = \frac{2}{16\pi^2} ( 6\lambda^2 A_\lambda + 6\kappa^2 A_\kappa 
	+ 3N_C \sum_{j=1}^{N_F}\lambda_{Q'_j}^2 A_{\lambda_{Q'_j}} ).   \label{Akappa}
\end{eqnarray}
The RGEs of the soft mass for the $Q'_i$ is given by
\begin{eqnarray}
	\frac{d m_{Q'_i}^2}{dt} = \frac{2}{16\pi^2}\left[ \lambda_{Q'_i}^2 
	(m_{Q'_i}^2 + m_{\bar{Q}'_i}^2 + |A_{\lambda_Q}|^2+ m_S^2) - 4C_2(r) g_H^2 |M_H|^2\right] .
\end{eqnarray}
The beta function for $m_S^2$ is also modified as
\begin{eqnarray}
	\frac{d m_S^2}{dt} = \left(\frac{d m_S^2}{dt}\right)_{\rm NMSSM} 
	+ \frac{2}{16\pi^2}\left[ N_C \sum_{i=1}^{N_F} \lambda_{Q'_i}^2(m_{Q'_i}^2 
	+ m_{\bar{Q}'_i}^2 + |A_{Q'_i}|^2 + m_S^2   ) \right].
\end{eqnarray}
Due to the largeness of $m_{Q'_i}^2$ and $m_{\bar{Q'}_i}^2$, 
$m_S^2$ can have sufficiently negative value at the EW scale.

  

\end{document}